\magnification=1200
\baselineskip=20 pt

\def\rc{r_c}
\def\hphi{\hat{\phi}}
\def\mphi{m_{\phi}}
\def\vphi{\langle \phi \rangle}

\def\a{\alpha}
\def\b{\beta}
\def\mh{m_h}
\def\l{\lambda}
\def\a{\alpha}
\def\b{\beta}
\def\dmu{\partial_{\mu}}
\def\dnu{\partial_{\nu}}

\hfill{hep-ph/0004128}

\centerline{\bf A quasi unitarity bound on the radion mass in the 
Randall-Sundrum model}

\vskip 1 true in

\centerline{\bf Uma Mahanta}
\centerline{\bf Mehta Research Institiute}
\centerline{\bf Chhatnag Road, Jhusi}
\centerline{\bf Allahabad-211019, India}

\vskip .4true in

\centerline{\bf Abstract}

In this paper we derive a quasi unitarity bound on the radion mass
($\mphi$) from the process $hh\rightarrow \phi\phi$. We find that
at sufficiently high energies (i.e. when $ s\gg m_h^2, \mphi^2 $) the J=0
partial wave amplitude for the above process violates unitarity if
$\mphi^2 > m_h^2+16 \pi \vphi^2 $. Combining this result with the unitarity
bound $m_h^2<{16\pi v\vphi\over 3}$ obtained from the process 
$hh\rightarrow h\phi$ we get an upper bound of $4\sqrt {\pi}\vphi
[1+{v\over 3 \vphi}]^{1\over 2}$ on the radion mass. This bound is however
valid only in the low energy effective theory where we can ignore 
the effects of the gravitational KK modes and the string/M theoretic 
excitations.

\vfill\eject

\centerline{\bf Introduction}

Recently several radical proposals based on extra dimensions have been
put forward to explain the large hierarchy between the Planck scale and 
the weak scale. Among them the Randall-Sundrum [RS] model [1] is most 
interesting because it propses a five dimensional world with a 
non-factorizable metric
$$ds^2= e^{-2k \rc | \theta |}\eta_{\mu\nu}dx^{\mu}dx^{\nu}-\rc^2 d\theta^2.
\eqno(1)$$

Here $\rc$ measures the size of the extra dimension which is an 
${S^1\over Z_2}$ orbifold. k is a mass parameter of the order of the
fundamental five dimensional Planck mass M.
 $x^{\mu}$ are the coordinates of the usual
four dimensional space time. $-\pi\le\theta\le \pi$ is the coordinate of
the extra dimension with $\theta$ and $-\theta$ identified. Two 
three-branes extending in the space time directions are placed at the 
orbifold fixed points $\theta=0$ and $\theta=\pi$. The 3 brane located
at $\theta =0$ is called the hidden (Planck) brane and the  3 brane
located at $\theta =\pi$ is called the visible brane.
 Randall and Sundrum
showed that any field with a fundamental mass parameter $m_0$
gets an effective four dimensional mass given by $m=m_0
e^{-k\pi\rc}$. Thus for $k\rc\approx 14$ the electro-weak scale
is generated from the fundamental Planck scale by the exponential
warp factor of the metric.

In the original proposal of Randall and Sundrum the compactification
radius $r_c$ was assumed to be determined by the vacuum expectation value
(vev) of scalar field $T(x)$. However the modulus field was massless 
and therefore its vev was not stabilized by some dynamics.
Goldberger and Wise showed [2] that by introducing a scalar field $\chi (x,
\theta )$ in the bulk with interactions localized on the two branes
it is possible to generate a potential for $T(x)$. They also showed that
the parameters of the potential can be adjusted to yield a minimum
at $k\rc \approx $ 14 without any extreme fine tuning.

The Randall-Sundrum model assumes that the Standard Model [SM]
fields are
 localized on the the visible brane at $\theta =\pi$. However the 
SM action is modified due to the exponential warp factor of the metric.
Fluctuations of the modulus field $T(x)$ about its vev $\rc$  
give rise to non-trivial couplings of the modulus field with the SM fields
on the visible brane. In fact
it was shown in Ref.[3] that linear fluctuations in the radion field 
$\phi$ couples to the SM fields on the visible brane through the Lagrangian
$$L_I={T^{\mu}_{\mu}\over \vphi}\hphi.\eqno(2)$$

Here $T^{\mu}_{\mu}$ is the trace of the energy-momentum tensor of the
SM fields localized on the visible brane. $\hphi$ is a small fluctuation
of the radion field from its vev and is given by $\phi =f e^{-k\pi T(x)}
=\vphi +\hphi$. $\vphi$ is the vev of $\phi$ and 
$f \approx \sqrt {24 M^3\over k}$ is a mass parameter
of the order of M.

In this paper we shall derive the couplings of a light stabilized radion
with the SM higgs scalar up to quadratic order in $\hphi$. We shall then
use these couplings to calculate the tree level transition amplitude
for the processes $hh\rightarrow \phi\phi$
and $hh\rightarrow h\phi$
 at very high energies i.e.
when $s\gg m_h^2, \mphi^2$. We shall however assume that $\sqrt{s}$ remains
much below the lowest gravitational KK mode and the string/M theoretic
excitation. This will enable us to neglect the contributions of the tower
of KK gravitons and stringy excitations to the transition amplitude
for the process $hh\rightarrow \phi\phi$. An exact computation of such 
contributions which involves summing over a tower of states is also quite 
difficult. In the Goldberger-Wise stabilzation mechanism the radion mass
( $\mphi \approx $ 10 Gev) indeed turn out to be much smaller than that
of the gravitational KK modes and string excitations which usually lie 
in the several Tev range. Hence an effective field theory involving
the SM particles and the light radion is quite realistic. By requiring
the J=0 partial wave amplitude for the process $hh\rightarrow \phi\phi$
in the context of this low energy effective theory  satisfies the unitarity
constraint we obtain an upper bound on $\mphi^2 -\mh^2$. Combining this 
result with the unitarity bound on $\mh$ obtained from the process
$hh\rightarrow h\phi$ we finally obtain an upper bound on the radion mass.

\centerline{\bf Radion couplings to the higgs scalar}

The couplings of the radion field to the SM higgs field localized on the
brane at $\theta =\pi$ is completely determined by general covariance. 
The action
for the SM higgs field in the Randall-Sundrum model is given by 

$$S=\int d^4x \sqrt{-g_v}[{1\over 2}g_v^{\mu\nu}\dmu h\dnu h-V(h)
 ].\eqno(3)$$

Here $g_v^{\mu\nu}$ is the induced metric on the visible brane. In the 
abscence of graviton fluctuations about the background metric
 it is given by $g_v^{\mu\nu}=
e^{2k\pi T(x)}\eta^{\mu\nu}= ({\phi\over f})^{-2}\eta^{\mu\nu}$
where $\eta^{\mu\nu}$ is the Minkowski metric. $\sqrt{-g_v}=\sqrt{-det
(g_v)}=e^{-4k\pi T(x)}=({\phi\over f})^4$.
The higgs potential  expanded about the classical minimum
 is given by
$$V(h)={\lambda\over 4} (h^4+4h^3v+4h^2v^2).\eqno(4)$$

In the above expression we have subtracted the vacuum energy from $V(h)$.
The mass of the higgs scalar can be determined from the above potential
and is given by $m_h^2=2\lambda v^2$. The couplings of the RS radion to the
higgs field is therefore given by 
$$S=\int d^4x[{1\over 2}\eta^{\mu\nu}({\phi\over f})^{ 2}\dmu h\dnu h
-{\lambda\over 4}({\phi\over f})^{ 4}(h^4+4h^3v+4h^2v^2)].\eqno(5)$$

Rescaling h and v according to $h\rightarrow {f\over \vphi} h$
and $v\rightarrow {f\over \vphi} v$ the
action becomes
$$S=\int d^4x[{1\over 2}\eta^{\mu\nu}({\phi\over \vphi})^2
 \dmu h\dnu h
-{\lambda\over 4}({\phi\over \vphi})^4   (h^4+4h^3v+4h^2v^2)].\eqno(6)$$

Expanding $\phi$ about its vev and keeping terms only up to quadratic order in 
$\hphi$ we get

$$\eqalignno{S &=\int d^4x [{1\over 2}\eta ^{\mu\nu}\dmu h\dnu h-
{\lambda\over 4}(h^4+4h^3v+4h^2v^2)]\cr
&+\int d^4x [\eta^{\mu\nu}\dmu h\dnu h- \lambda (h^4+4h^3v+4h^2v^2)]
{\hphi\over \vphi}\cr
&+\int d^4x {1\over 2}[\eta^{\mu\nu}\partial_{\mu}h \partial_{\nu}h
-3\lambda (h^4+4h^3v+4h^2v^2)]{\hat {\phi}^2\over \vphi^2 }+
..&(7)\cr}$$

The dots stand for the couplings of cubic and higher order fluctuations
in the radion field to the higgs scalar. These couplings will not
be needed for the processes under consideration in this paper.

\centerline{\bf Unitarity bound from the process $hh\rightarrow\phi\phi$}

We shall now derive the unitarity bound that follows from the process
$hh\rightarrow \phi\phi$ in the context of the low energy effective
field theory where we can legitimately neglect the contributions
from gravitational KK modes and string excitations. At tree level the
process $hh\rightarrow \phi\phi$ gets contributions from three distinct
Feynman diagrams. Let $M_1$ and $M_2$ be transition amplitudes for
t and u channel higgs exchanges. $M_3$ be the transition amplitude for 
the contact interaction diagram. It can then be shown that

$$M_1={1\over 2\vphi^2}[(t-\mh^2 )-2(2\mh^2+\mphi^2)+{(2\mh^2+\mphi^2)^2\over
(t-\mh^2 )}].\eqno(8)$$

$$M_2={1\over 2\vphi^2}[(u-\mh^2 )-2(2\mh^2+\mphi^2)+{(2\mh^2+\mphi^2)^2\over
(u-\mh^2 )}].\eqno(8)$$

and 

$$M_3={1\over 2\vphi^2}[s+10\mh^2 )].\eqno(9)$$

At sufficiently high energies i.e. when $s\gg \mh^2, \mphi^2$ but
$s<m^2_{0g}, m^2_{0s}$ ($m_{0g}$ and $m_{0s}$ being the masses of the
lightest gravitational KK mode and string excitation respectively)
the total transition amplitude exhibits the following asymptotic
behaviour

$$\eqalignno{M &=M_1+M_2+M_3={1\over 2\vphi^2}[2(\mh^2-\mphi^2)\cr
&+(2\mh^2+\mphi^2)^2({1\over (t-\mh^2)}+{1\over (u-\mh^2)})]\cr
&\approx {(\mh^2-\mphi^2)\over \vphi^2} +O({1\over s})&(10)\cr}$$

Following Lee, Quigg and Thacker [4] if
we require that the J=0 partial wave amplitude must satisfy the 
constraint $|a_0|<1$ [5] we get the unitarity bound $|\mphi^2-\mh^2|<16\pi
\vphi^2$.

\vfill\eject

\centerline{\bf Unitarity bound on $\mh$ from the process $hh\rightarrow
h\phi$}

The transition amplitude for the process $hh\rightarrow h\phi$ gets 
contributions only from radion couplings in the linear approximation.
Since the gravitational KK modes do not contribute to the process
$hh\rightarrow h\phi$ the bound on $\mh$ that we shall present below will
be valid more generally in the complete model of Randall and Sundrum.
Let $M_1$, $M_2$ and $M_3$ denote the transition amplitudes due to
s, t and u channel higgs exchanges for the process $hh\rightarrow h\phi$.
$M_4$ be the transition amplitude for the conatct interaction diagram.
We then find that at very high energies to leading  order in ${\mh^2\over s}$
and ${\mphi^2\over s}$

$$M_1={-6\l v\over\vphi}[1-{(2\mh^2+\mphi^2)\over (s-\mh^2)}]\approx
{-6\l v\over \vphi}(1-{2\mh^2+\mphi^2\over s}).\eqno(11)$$

$$M_2={-6\l v\over\vphi}[1-{(2\mh^2+\mphi^2)\over (t-\mh^2)}]\approx
{-6\l v\over \vphi}(1+{2\over s}{2\mh^2+\mphi^2\over (\a-\b x)}).\eqno(11)$$

$$M_3={-6\l v\over\vphi}[1-{(2\mh^2+\mphi^2)\over (u-\mh^2)}]\approx
{-6\l v\over \vphi}(1+{2\over s}{2\mh^2+\mphi^2\over (\a+\b x)}).\eqno(12)$$

and 

$$M_4={24\l v\vphi }.\eqno(13)$$

where $\a =1-{\mh^2+\mphi^2\over s}$ and $\b =1-{3\mh^2+\mphi^2\over s}$.
$x= \cos\theta$ where $\theta$ is the angle between the outgoing h
and one of the incoming h. At sufficiently high energies i.e. for
$s\gg \mh^2, \mphi^2$ the J=0 partial wave amplitude exhibits the 
following asymptotic behaviour

$$a_0\approx {3\l v\over 8\pi\vphi}[1+{2\mh^2+\mphi^2\over s}-
{2\over \b}{2\mh^2+\mphi^2\over s}\ln {\a+\b\over \a-\b }].\eqno(14)$$

The unitarity constraint then gives rise to the bound $\mh^2<{16\pi\over 3}
\vphi v$. We would like to note first that in the purely SM a similar
unitarity bound on $\mh$ can be derived from the process $hh\rightarrow
hh$. Second the radion mass $\mphi$ decouples from the asymptotic
form of the transition amplitude for the process $hh\rightarrow h\phi$.
This causes the unitarity bound on $\mh^2$ derived in this paper
to depend linearly on v in contrast to its quadratic 
dependence on v in the SM.

\centerline{\bf Results and Conclusion}

Combining the unitarity bound on $\mh$  derived from the process 
$hh\rightarrow h\phi$ with the result $|\mphi^2-\mh^2| <16\pi \vphi^2$
we obtain $\mphi<4 \sqrt {\pi}\vphi [1+{v\over 3\vphi}]^{1\over 2}$.
Since the process $hh\rightarrow \phi\phi$ can receive contributions
from gravitational KK modes and string excitations and we have ignored
such contributions the above bound on $\mphi$ is valid only in the low 
energy effective theory involving the SM particles and the light radion.
It is in this sense that we have called the bound a quasi unitarity
bound. The bound on $\mh$ is however valid more generally in the context
of the complete Randall-Sundrum model. Note that the results obtained in 
this paper should be valid irrespective of the value of $\vphi$. Thus
for GUT size compactification where $\vphi \approx 10^{15}$
Gev we would expect unitarity violation in the process $hh\rightarrow
h\phi$  to occur if 
$\mh >10^9$ Gev. However for the same type of compactification
unitarity violation would occur in the process $hh\rightarrow \phi\phi$
if $\mphi >5\times 10^{15}$ Gev. Hence for $\vphi \gg v$ the unitarity
bound on $\mphi$ is determined mainly by $\vphi$. On the other hand
if $\vphi\ll v$ the uniatity bounds on both $\mh$ and $\mphi$ will
be much small compared to the EW symmetry breaking scale.

\centerline{\bf References}

\item{1.} L. Randall and R. Sundrum, Phys. Rev. Lett 83, 3370 (1999).

\item{2.} W. D. Goldberger and M. B. Wise, Phys. Rev. Lett. 83, 4922 (1999).

\item{3.} C. Csaki, M. Graesser, L. Randall and J. Terning, hep-ph/9911406;
W. D. Goldberger and M. B. Wise, hep-ph/ 9911457.

\item{4.} B. W. Lee, C. Quigg and H. B. Thacker, Phys. Rev. D, 16, 1519 (1977);
D. Dicus and V. Mathur, Phys. Rev. D 7, 3111 (1973).

\item{5.} S. Dawson and and S. Willenbrock, Phys. Rev. Lett. 62, 1232 (1989);
W. Marciano, G. Valencia and S. Willenbrock, Phys. Rev. D 40, 1725 (1989);
L. Durand, J. Johnson and J. Lopez, Phys. Rev. Lett. 64, 1215 (1990).

\end